\newcommand{\bld}[1]{\mbox{\boldmath{$#1$}}}

\documentstyle[preprint,aps,12pt,floats]{revtex}
\begin{document}
\draft
\tighten
\renewcommand{\floatpagefraction}{0.7}



\begin{titlepage}
\hfill \parbox{5cm}{\raggedleft TTP96--50\\ physics/9611004\\November 1996}
\vspace*{1.5cm}
\begin{center}{\LARGE Eight--component differential equation for leptonium}\\
\vspace{1cm}
\large{Ruth H\"ackl, Viktor Hund and Hartmut Pilkuhn}\\
\vspace{0.5cm}
{\em Institut f\"ur Theoretische Teilchenphysik,\\ Universit\"at Karlsruhe,\\ 
D-76128 Karlsruhe, Germany}
\end{center}

\vspace{2\baselineskip}
\begin{abstract}{
It is shown that the potential for lepton-antilepton bound 
states (leptonium) is
the Fourier transform of the first Born approximation to the QED scattering
amplitude in an 8-component equation, while 16-component equations are
excluded. The Fourier transform is exact at all {\sl cms} energies $ - \infty <
E < \infty$; the resulting atomic spectrum is explicitly CPT-invariant.
\begin{center}{ PACS number: 03.65.Pm} \end{center}
}\end{abstract}
\vspace{4cm}
-----------\\
{\scriptsize \begin{tabular}{ll} 
Fax: & +49-721-370726 \\ 
Internet: & 
Ruth.Haeckl@physik.uni-karlsruhe.de\\
 &  vh@ttpux1.physik.uni-karlsruhe.de\\
 & hp@ttpux2.physik.uni-karlsruhe.de \end{tabular} }
\end{titlepage}

\setcounter{page}{2}

\section{Introduction}\label{einl}

In muonium ${\mbox e}^{-}\mu^{+}$, the mass 
$m_{2}$ of the heavier particle is so large
that the kinetic energy $p_{2}^{2}/2m_{2}$ can be added as a recoil correction
to the electron's Dirac Hamiltonian. The resulting Dirac equation has 8
components, with the Pauli matrices $\bld{\sigma}_{2}$ in the hyperfine
operator \cite{grotch,sapir}. However, a similar 
8--component equation exists which avoids
the expansion in terms 
of $1/m_{2}$ \cite{pil1,mal}. Despite its asymmetric form, it
reproduces the energy levels also for positronium ${\mbox e}^{-}{\mbox e}^{+}$
to order $\alpha^{4}$ \cite{pil2}. 
Unfortunately, its complicated derivation via
Breit operators introduced a low-energy approximation,
 which is removed in the present paper. The new
interaction is simply the Fourier transform of the QED Born term in the {\sl
cms}, valid at all  energies, $-\infty < E < \infty$ (by crossing
symmetry, $E < 0$ for ${\mbox e}^{-} \mu^{+}$-scattering refers to $E > 0$ for
${\mbox e}^{+} \mu^{-}$-scattering). The equation is written in the {\sl cms},
${\bf p} = {\bf p}_{1} = - {\bf p}_{2}$, in units $\hbar = c = 1$:
\begin{eqnarray}\label{eins}
\lefteqn{
\frac{1}{2} E^{2} \psi = ( K_{0} + K_{I} ) \psi \, ,\,\,\, K_{0} = E {\bf p}
\bld{\alpha} + \frac{1}{2} ( m_{2} + \beta m_{1} ) ^{2}\, , }\\
\label{zwei} & & 
\hspace{0.3cm} 
\bld{\alpha} = \gamma_{5} \bld{\sigma}_{1}\, , \qquad \gamma_{5} = \left(
\begin{array}{cc}
0 & 1 \\
1 & 0 \end{array} \right)\, , \qquad \beta = \left( \begin{array}{cr}
1 & 0 \\
0 & -1 \end{array} \right)\, .
\end{eqnarray}
The Pauli matrices $\bld{\sigma}_{2}$ appear only in the interaction operator
$K_{I}$. The proof that the equation with $K_{I} =0$ describes two free Dirac
particles instead of one is given in section \ref{free}. The construction of
$K_{I}$ from the Born approximation is done in section \ref{born}. 
Orthogonality
relations are derived in section \ref{a4}.

Before going into details, we remind the reader that in a 16-component version,
\begin{equation}\label{drei}
E \psi = (H_{1} + H_{2} + H_{12} ) \psi \, , \quad H_{i}= {\bf p}_{i}
\bld{\alpha}_{i} + m_{i} \beta_{i}\, ,
\end{equation}
the energy levels to order $\alpha^{4}$ are not reproduced when $H_{12}$ is the
Fourier transform of the Born approximation, $H_{12} = V ( 1 -
\bld{\alpha}_{1}\bld{\alpha}_{2} )\,\,$ ($V= - \alpha / r \, , \,\, \alpha =
e^{2}$).  This interaction was improved by Breit \cite{breit} and 
was later amended
by energy-transfer \cite{bethe,grant} and by  
positive-energy projectors \cite{lind}. 
The method is now quite successful in atomic theory, where the nucleus
produces an external potential for the electrons \cite{grant}.
In the previous derivation of
(\ref{eins}) via Breit operators \cite{pil1,mal}, a simpler 
prescription was used which
sets the (large) squares of Breit operators equal to zero. However, the
connection with perturbative QED was then incomplete. In the present
derivation, the connection with the Born series is straightforward. With loop
graphs included, all $\alpha^{6}$-effects in binding energies should appear,
for arbitrary values of $m_{1}/m_{2}$.

For $V=0$, (\ref{eins}) and (\ref{drei}) are equivalent. In the first Born
approximation, $K_{I}$ of (\ref{eins}) and $H_{12}$ of (\ref{drei}) are linear
in $V$. When (\ref{drei}) is reduced to 8 components, quadratic terms appear
which contradict the leptonium energy levels to order $\alpha^{4}$. Some of
these terms were eliminated by a canonical transformation, the rest by
projectors. 

To understand the significance  of the 8-component theory, 
one may note that 8 is the
minimum number of components for a Lorentz- and parity invariant theory.
An irreducible representation of the Lorentz group for a single lepton requires
only two components, for example the right-handed $\psi_{R}$ in a chiral
basis. It is the parity transformation, $\psi\,  ' 
( -{\bf r}) = \beta \psi ( {\bf
r} )$, which exchanges $\psi_{R}$ with the left-handed components
$\psi_{L}$ (in the chiral or Weyl basis, $\gamma_{5}$ and $\beta$ of
(\ref{zwei}) exchange their roles). For two leptons, the necessary $2 \times 2
= 4$ components are doubled by the matrix $\beta$ of (\ref{zwei}), which
corresponds to $\beta_{1}\beta_{2}$ in a 16-component formalism. The separate
matrices $\beta_{1}$ and $\beta_{2}$ of the 16-component formalism provide
another 
unnecessary doubling.

We end this section by casting (\ref{eins}) into a more familiar form. We
divide it by $E$ and define the reduced mass $\mu$ and reduced energy
$\epsilon$:
\begin{eqnarray}\label{vier}
\lefteqn{
\mu = m_{1} m_{2} / E \, ,\quad \epsilon = ( E^{2} - m_{1}^{2} - m_{2}^{2} )
/ 2 E \, , } \\
\label{fuenf} & & 
\hspace{1cm}
\epsilon \psi = ( {\bf p} \bld{\alpha} + \mu \beta + K_{I} /  E ) \psi \, . 
\end{eqnarray}
We also anticipate the result (\ref{40}) for $K_{I}$, which gives:
\begin{equation}
\label{sechs} 
K_{I} / E = V - i \gamma_{5} V {\bf p} ( \bld{\sigma}_{1} \times
\bld{\sigma}_{2} ) / E\, .
\end{equation}
The difference to an effective Dirac equation is obviously restricted to the
hyperfine operator $ - i V {\bf p} ( \bld{\sigma}_{1} \times \bld{\sigma}_{2} )
/ E$. As the {\sl cms} energy $E$ is close to $m_{1} + m_{2}$ for ordinary
bound states, the replacement of the familiar $1/m_{2}$ by $1/E$ is a recoil
correction. The essential new feature is that the operator $V\! {\bf p}$ is not
hermitian. An 8-component equation with a hermitian operator does not reproduce
the positronium bound states.

Anomalous magnetic moments are neglected for the time being, so that the
present formulation does not cover atomic hydrogen. The annihilation graph in
the $^{3} S_{1}$ state of positronium is also neglected.
The hyperfine operator will be further simplified in appendix \ref{appa}.
Its connection with the previous approximate operator \cite{mal,pil2} is given
in appendix \ref{appb}. Appendix \ref{appc} contains a spin summation, which is
included for demonstration of covariance.

\section{Eight-component spinors for two  leptons}\label{free}

Let $u_{ig}$ and $u_{if}$ denote the large and small components of the Dirac
spinor $u$ of a particle $i$ in an  orbital:
\begin{equation}\label{7n}
{\bf p}_{i}\bld{\sigma}_{i} \, u_{ig} = ( \pi _{i}^{0} + m_{i} ) \, u_{if}\, ,
\quad {\bf p}_{i} \bld{\sigma}_{i} 
\, u_{if} = ( \pi_{i}^{0} - m_{i} ) \, u_{ig}
\, , \quad \pi_{i}^{0} = E_{i} - V_{e} ( {\bf r}_{i} ) \, .
\end{equation}
Here $V_{e}$ is a possible external potential, as in He-atoms.
Leptonium has $V_{e} = 0$, of course.
The chiral operator $\gamma_{5i}$ exchanges $u_{ig}$ with $u_{if}$. Out of the
4 products $u_{1g}u_{2g} \dots u_{1f}u_{2f}$, we keep in this section only the
two combinations of total chirality $+1$ (the eigenvalue of
$\gamma_{51}\gamma_{52}$): 
\begin{eqnarray}\label{9n}
\frac{1}{\sqrt{2}} ( u_{1g} u_{2g} + u_{1f} u_{2f} )  & \equiv  & u_{dg}\, ,
\nonumber \\ 
\frac{1}{\sqrt{2}} ( u_{1f} u_{2g} + u_{1g} u_{2f} ) & \equiv & u_{df} \, .
\end{eqnarray}
Inspection shows that these combinations satisfy the following equations:
\begin{eqnarray}\label{10n}
( m_{2} {\bf p}_{1} \bld{\sigma}_{1} + m_{1} {\bf p}_{2} 
\bld{\sigma}_{2} ) \, u_{dg} &
= & ( m_{2} \pi_{1}^{0} + m_{1} \pi_{2}^{0} ) \, u_{df}\, , \nonumber \\
( m_{2} {\bf p}_{1} \bld{\sigma}_{1} - m_{1} {\bf p}_{2} 
\bld{\sigma}_{2} ) \, u_{df}&
 = &
( m_{2} \pi_{1}^{0} - m_{1} \pi_{2}^{0} ) \, u_{dg} \, .
\end{eqnarray}
Apart from the $m_{1}$-terms, these equations have the same structure as a
single-particle Dirac equation. We therefore introduce an $8\times 8$-matrix
$\beta$ (which has the same eigenvalues as $\beta_{1}\beta_{2}$) and an
anticommuting matrix $\gamma_{5}$, $\gamma_{5} \beta = - \beta \gamma_{5}$, and
combine $u_{dg}$ and $u_{df}$ into one 8-component double spinor $u_{d}$:
\begin{equation}\label{11n}
u_{d}  = \left( 
\begin{array}{c} u_{dg} \\ u_{df} \end{array} \right) \, , \quad
\gamma_{5} ( m_{2} {\bf p}_{1} \bld{\sigma}_{1} + \beta m_{1} {\bf p}_{2}
\bld{\sigma}_{2} )\,  u_{d} = (m_{2} \pi_{1}^{0} - \beta m_{1} \pi_{2}^{0} )\, 
u_{d}
\, .
\end{equation}
$\gamma_{5}$ and $\beta$ have already been given in (\ref{zwei}), understanding
that the number $1$ is a $4\times 4$-matrix in spin space. For $m_{1} = m_{2}$,
(\ref{11n}) should also describe two non-interacting electrons in a helium
atom. Their Born scattering amplitude for $V_{e} \not= 0$ will be constructed
in (\ref{36n}) below, but our construction of the interaction operator in the
differential equation neglects $V_{e}$.
Turning now to the free two-body problem, we set ${\bf p}_{1} = - {\bf p}_{2} =
{\bf p}\,$,
\begin{equation}\label{12n}
\pi_{1}^{0} = E_{1} = \frac{1}{2E} ( E^{2} + m_{1}^{2} - m_{2}^{2} )\, , \quad
\pi_{2}^{0} = E_{2} = \frac{1}{2E} ( E^{2} - m_{1}^{2} + m_{2}^{2} ) \, .
\end{equation}
A factor $m_{2} - \beta m_{1}$ can then be divided off from the right-hand side
of (\ref{11n}), the result being:
\begin{eqnarray}\label{13n}
\lefteqn{
( \bld{\alpha}^{c} {\bf p} + \mu \beta - \epsilon ) \, u_{d} = 0 \, ,
}\\ \label{14n}
& & 
\hspace{-2.5cm} 
\bld{\alpha}^{c} = \gamma_{5} \bld{\sigma}_{1}^{c} = \gamma_{5} (m_{2}
\bld{\sigma}_{1} - \beta m_{1} \bld{\sigma}_{2} ) / (m_{2} + \beta m_{1} ) \, .
\end{eqnarray}
Including the space-dependence, the free double-spinor wave function in the
{\sl cms}  is:
\begin{equation}\label{15n}
\psi_{d} ( {\bf r} ) = u_{d}\, e^{i{\bf k}_{1} {\bf r}_{1}} e^{i {\bf
k}_{2}{\bf r}_{2}} = u_{d} \, e^{i{\bf kr}}\, , \quad {\bf r} = {\bf r}_{1} -
{\bf r}_{2} \, .
\end{equation}
It differs from the corresponding free-particle solution of (\ref{eins}), which
has $\bld{\alpha}^{c}$ replaced by $\bld{\alpha}$. This difference may be one
reason why (\ref{eins}) has not been discovered 60 years ago. During these 60
years, many different formalisms have been developed. 
Bethe and Salpeter advocated the use
of four-dimensional integral 
equations (with a relative time as fourth integration
variable), again with 16 components. Although the relative-time concept turned
out to be useless, one learned to find bound states from integral equations in
momentum space, now in three dimensions \cite{sapir}. 
This method avoids the Fourier
transformation. Having elaborated these momentum space methods, one may be
unwilling to return to differential equations, particularly if these require
such strange matrices as $\bld{\sigma}_{1}^{c}$. On the other hand, the
advantage
of the present formulation survives also in 8-component integral
equations in momentum space. 
It
could become
essential also in quarkonium models.

To establish the connection between (\ref{eins}) and (\ref{13n}), we first
define two mass operators:
\begin{equation}\label{16n}
m_{\pm} = m_{2} \pm \beta m_{1} \, , \quad m_{+}m_{-} = m_{2}^{2} - m_{1}^{2}
\, , \quad m_{\pm}^{2} = m_{1}^{2} + m_{2}^{2} \pm 2m_{1} m_{2} \beta \, .
\end{equation}
The expression for $m_{+}^{2}$ has already been used in (\ref{fuenf}). From
$\gamma_{5} \beta = - \beta \gamma_{5}$, one finds:
\begin{equation}\label{17n}
\gamma_{5} m_{+} = m_{-} \gamma_{5} \, .
\end{equation}
The Dirac spin operators are $\gamma_{5}\bld{\sigma}_{1}$ and $\gamma_{5}
\bld{\sigma}_{2}$. With the algebra (\ref{16n}), (\ref{17n}), one easily
verifies:
\begin{equation}\label{18n}
\left( \bld{\alpha}^{c} \right) ^{2} = \left( \gamma_{5} \bld{\sigma}_{1}^{c}
\right) ^{2} = (m_{2} \bld{\sigma}_{1} + \beta m_{1} \bld{\sigma}_{2} ) ( m_{2}
\bld{\sigma}_{1} - \beta m_{1} \bld{\sigma}_{2} )/ m_{+} m_{-} = 3 \, .
\end{equation}
Comparing this with $\bld{\alpha}^{2} = 3$, one sees that there should exist a
transformation from $\bld{\alpha}^{c}$ to $\bld{\alpha}$. It was first found in
an explicit decomposition of $u_{d}$ into $u_{dg}$ and $u_{df}$, and of the
spin states $\chi_{12}$ into $\chi_{s}$ (singlet) and $\chi_{t}$ (triplet,
eq. (5.14) in \cite{pil1}). Its compact Dirac form is:
\begin{equation}\label{19n}
\psi_{d} = c \psi \, , \quad c = (m_{+}m_{-} ) ^{-1/2} \left\lbrack m_{+} - 2
m_{1} \beta P_{s} \right\rbrack \, . 
\end{equation}
$P_{s}$ is the projector on the singlet spin state $\chi_{s}$. In the
following, the corresponding triplet projectors on the three states $\chi_{t}$
will also be needed:
\begin{equation}\label{20n}
P_{s} = \frac{1}{4} ( 1 - \bld{\sigma}_{1} \bld{\sigma}_{2} ) \, ,\quad P_{t} =
\frac{1}{4} ( 3 + \bld{\sigma}_{1} \bld{\sigma}_{2} ) \, .
\end{equation}
We also define  combinations of Pauli matrices,
\begin{equation}\label{21n}
\bld{\sigma} = \bld{\sigma}_{1} + \bld{\sigma}_{2} \, , \quad \Delta
\bld{\sigma} = \bld{\sigma}_{1} - \bld{\sigma}_{2} \, , \quad \bld{\sigma}
^{\times}  = \bld{\sigma}_{1} \times \bld{\sigma}_{2} \, ,
\end{equation}
which have the following products with $P_{s}$ and $P_{t}$:
\begin{equation}\label{22n}
\bld{\sigma} P_{s} = P_{s} \bld{\sigma} = 0 \, , \quad \Delta 
\bld{\sigma} P_{t} = P_{s}
\Delta \bld{\sigma} \, , \quad \Delta \bld{\sigma} P_{s} = P_{t} \Delta
\bld{\sigma} \, , \quad \bld{\sigma}^{\times} P_{s} = P_{t}
\bld{\sigma}^{\times} \, .
\end{equation}
$c$ is not unitary, its inverse being:
\begin{equation}\label{23n}
c^{-1} = (m_{+}m_{-} ) ^{-1/2} (m_{+} - 2 m_{1} \beta P_{t} ) \, ,
\end{equation}
which is checked by using $P_{s} + P_{t} = 1, \, P_{s}P_{t} = 0$. An important
property of $c$ is:
\begin{equation}\label{24n}
c^{-1} \gamma_{5} = \gamma_{5} c \, .
\end{equation}
By parity invariance, the operators (\ref{21n}) are always accompanied by one
factor $\gamma_{5}$, which in view of (\ref{24n}) replaces $c^{-1}$ by $c$:
\begin{equation}\label{25n}
c\, \bld{\sigma} \, c = \bld{\sigma} \, m_{+} /m_{-}\, , \quad c\, \Delta
\bld{\sigma} \, c = \Delta \bld{\sigma}\, , \quad c \, \bld{\sigma}^{\times} \,
c = \bld{\sigma} ^{\times}\, .
\end{equation}
Expressing $\bld{\sigma}_{1}$ as $\frac{1}{2} ( \bld{\sigma} + \Delta
\bld{\sigma} )$, one finds for the combination (\ref{14n}):
\begin{equation}\label{26n}
c\, \bld{\sigma}_{1}^{c} \, c = \bld{\sigma}_{1} \, .
\end{equation}
Thus the transformation (\ref{19n}) leads from (\ref{13n}) to the free equation
(\ref{eins}). 
Other forms of that equation are generated by additional transformations $d$
satisfying:
\begin{equation}
\beta d = d \beta \, \quad d^{-1} \gamma_{5} =\gamma_{5} d \, .
\end{equation}
%

\section{The Born approximation and its Fourier transform}\label{born}

The Lorentz-invariant $T$-matrix for lepton-antilepton scattering 
from initial orbitals 1,2 into final orbitals $1',2'$
has the
following Born approximation:
\begin{equation}\label{28n}
T/4\pi = \alpha {\bar u}'_{1} \gamma^{\mu} u_{1} {\bar u}'_{2} \gamma_{\mu}
u_{2} / t = \alpha u_{1}'^{\dagger} u_{2}'^{\dagger} ( 1 -
\bld{\alpha}_{1}\bld{\alpha}_{2} ) u_{1} u_{2} / t\, ,
\end{equation}
with $\bld{\gamma}_{i} = \beta_{i} \bld{\alpha}_{i}$, and
$t= q^{\mu} q_{\mu}= q_{0}^{2} - {\bf q}^{2}$ 
being the square of the 4-momentum transfer.
 In the {\sl cms}, the
arguments of the free Dirac  spinors are ${\bf k}$ and $-{\bf k}$ in the
initial state and ${\bf k}'$ and $-{\bf k}'$ in the final state, and ${\bf q} =
{\bf k} - {\bf k}' \, , \,\,  q_{0} = 0\, .$
Remembering
\begin{equation}\label{29n}
u_{1}'^{\dagger} u_{1} u_{2}'^{\dagger} u_{2} = \left( u_{1g}'^{\dagger} u_{1g}
+ u_{1f}'^{\dagger} u_{1f} \right) \left( u_{2g}'^{\dagger} u_{2g} +
u_{2f}'^{\dagger} u_{2f}\right)
\end{equation}
etc., one sees that $T$ cannot be written as a bilinear in $u_{d}$ and
$u_{d}'^{\dagger}$. One also needs the states of total chirality $-1$, which
will be called $w_{dg}$ and $w_{df}$:
\begin{eqnarray}\label{30n}
w_{dg} & = & \frac{1}{\sqrt{2}} ( u_{1g} u_{2g} - u_{1f} u_{2f} ) \, ,
\nonumber \\
w_{df} & = & \frac{1}{\sqrt{2}} ( u_{1f} u_{2g} - u_{1g} u_{2f} 
)\, , \\
\label{31n}
T/4\pi  =  \alpha \left\lbrack u_{d}'^{\dagger} ( 1\right.
 & - & \left. \bld{\sigma}_{1}
\bld{\sigma}_{2} )u_{d} + w_{d}'^{\dagger} ( 1 + \bld{\sigma}_{1}
\bld{\sigma}_{2} ) w_{d} \right\rbrack / t \, . 
\end{eqnarray}
This form still has its 16-component character, as the replacement of
$u_{1}u_{2}$ by $u_{d}$ and $w_{d}$ is just a unitary transformation. However,
in addition to the separate equations for $u_{d}$ and $w_{d}$ (the equation for
$w_{d}$ has $\bld{\sigma}_{2}$ replaced by $-\bld{\sigma}_{2}$)
there exist also coupled equations, with $\pi^{0} = \pi_{1}^{0} + \pi_{2}^{0}$
and $p_{\pm} = {\bf p}_{1} \bld{\sigma}_{1} \pm {\bf p}_{2} \bld{\sigma}_{2}$:
\begin{equation}\label{32n}
w_{d} = m_{+}^{-1} ( \pi^{0}  - \gamma_{5}   p_{+} )u_{d}\, ,
\quad u_{d} = m_{+}^{-1} ( \pi^{0} - \gamma_{5} p_{-})w_{d}\, ,
\end{equation}
which can be verified explicitly from (\ref{9n}) and (\ref{30n}).
By means of (\ref{32n}), $T$
can be written in terms of a single $8\times8$-matrix $M$,
\begin{eqnarray}
\label{33n}
\lefteqn{ \hspace{3.5cm}
T/ 4 \pi  =   \alpha \, w_{d}'^{\dagger}\, M\,  u_{d}\,  / t \, ,
} \\
\label{34n} & & 
M  =  ( \pi^{0} - \gamma_{5} p_{-}' ) m_{+}^{-1} ( 1 -
\bld{\sigma}_{1} \bld{\sigma}_{2} ) + ( 1 + \bld{\sigma}_{1} \bld{\sigma}_{2} )
m_{+}^{-1} ( \pi^{0} - \gamma_{5} p_{+} ) \, .
\end{eqnarray}
The operators proportional to $\pi^{0}$ combine into $2\pi^{0}$, 
while the operators 
containing $p_{-}'$ combine as follows:
\begin{equation}\label{35n}
p_{-}' \, ( 1 - \bld{\sigma}_{1} \bld{\sigma}_{2} ) =  {\bf p} 
( \Delta \bld{\sigma} + i \bld{\sigma}^{\times} ) = p_{-} ( 1 -
\bld{\sigma}_{1} \bld{\sigma}_{2} ) \, .
\end{equation}
%
The total momentum ${\bf p} = {\bf p}_{1} + {\bf p}_{2} = {\bf p}_{1}' + {\bf
p}_{2}' $ commutes with $V$ and vanishes
in the {\sl cms}. There, $M$ reduces to:
\begin{equation}\label{36n}
M= m_{+}^{-1} \left\lbrack 2 E - \gamma_{5}
(1+\bld{\sigma}_{1}\bld{\sigma}_{2}){\bf k} \Delta \bld{\sigma} \right\rbrack =
2 m_{+}^{-1} ( E - i \gamma_{5} {\bf k} \bld{\sigma}^{\times} ) \, .
\end{equation}
An equivalent form of $T$ follows from the elimination of 
$w_{d}'^{\dagger}$ and
$u_{d}$ in (\ref{31n}),
\begin{eqnarray}\label{37n}
\lefteqn{ \hspace{3.5cm}
T/4 \pi =  \alpha \, u_{d}'^{\dagger}\,  M^{\dagger}\,  w_{d}/t \, ,
}
\\ 
\label{38n}
& & 
M^{\dagger} = \left\lbrack 2 E - \gamma_{5} {\bf k}' \Delta \bld{\sigma} ( 1 +
\bld{\sigma}_{1} \bld{\sigma}_{2} ) \right\rbrack m_{+}^{-1} = 2 ( E + i
\gamma_{5} {\bf k}' \bld{\sigma}^{\times} ) m_{+}^{-1} \, .
\end{eqnarray}
This suggests the definition of a second double-spinor wave-function as
follows:
\begin{equation}\label{39n}
\chi_{d}({\bf r}) = e^{i {\bf kr}}\,  m_{+} \, w_{d}\, .
\end{equation}
When $T$ is expressed in terms of $m_{+} w_{d}$ and $w_{d}'^{\dagger}m_{+}$,
the factor $m_{+}^{-1}$ vanishes both in (\ref{36n}) and (\ref{38n}). The
interaction in coordinate space follows as:
\begin{eqnarray}\label{40n}
\lefteqn{ - 4 \pi\,  \alpha / {\bf q}^{2} = \int d^{3} 
r e^{-i{\bf k}'{\bf r}}\,  V\,  e^{i{\bf kr}}\, ,
\,\,\, - 4 \pi \, \alpha\,  {\bf k} / {\bf q}^{2} = \int d^{3} r
 e^{-i{\bf k}' {\bf
r} } \, V
{\bf p}\, e^{i{\bf kr} }\, ,} \\ \label{41n}
& &\hspace{3cm}  -4\pi\,  \alpha\, {\bf k}' / {\bf q}^{2} = \int d^{3} 
r e^{-i{\bf k}' {\bf r}}\, {\bf p} 
V \, e^{i{\bf kr}} \, .
\end{eqnarray}
It produces the operator
\begin{equation}\label{40}
K_{I} = V ( E - i {\bf p} \bld{\sigma}^{\times}\gamma_{5} )\, , \quad
K_{I}^{\dagger} = (E + i {\bf p} \bld{\sigma}^{\times} \gamma_{5} ) V\, ,
\end{equation}
to be used in (\ref{eins}) and
%
%
in the corresponding equation for the wave-function $\chi ({\bf r}) = e^{i{\bf
kr} } w$:
\begin{equation}\label{43n}
\frac{1}{2} E^{2} \chi = (K_{0} + K_{I}^{\dagger} ) \, \chi \, .
\end{equation}
Although $K_{I}$ is not hermitian, $K_{I}$ and $K_{I}^{\dagger}$ give
equivalent equations, such that the bound state energies may be real.

\section{Orthogonality relations and concluding remarks}\label{a4}

When $E$ is replaced by $m= m_{1} + m_{2}$ in the hyperfine operator, 
(\ref{fuenf}) and the corresponding equation (\ref{43n}) assume Hamiltonian
forms:
\begin{equation}\label{44n}
\epsilon \psi = H \psi \, ,  \quad \epsilon \chi = H^{\dagger} \chi \, .
\end{equation}
Taking the hermitian adjoint of the second equation at reduced energy $\epsilon
'$ and integrating over ${\bf r}$, one obtains
\begin{equation}\label{45n}
(\epsilon - \epsilon ' ) \int \chi '^{\dagger} \psi = \int \chi '^{\dagger} (
H - H ) \psi = 0\, .
\end{equation}
Thus the non-hermiticity of $H$ is harmless. But in the exact expression
(\ref{fuenf}), the hyperfine operator will 
remain in 
the orthogonality
relations.

One may also cast (\ref{eins}) into a strictly Hamiltonian form by introducing
a secondary 8-component spinor $\psi_{s}$:
\begin{equation}\label{46n}
( E - 2 \bld{\alpha} {\bf p} ) \psi = \psi_{s}\, , \,\,\, E \psi_{s} = (
m_{+}^{2} + 2 K_{I} ) \psi \, .
\end{equation}
This method is known from the relativistic treatment of spinless particles (for
example from the Klein-Gordon equation). 

For $V= - \alpha / r$, orthogonality
relations are most elegantly derived in a dimensionless scaled variable,
\begin{equation}\label{47n}
{\tilde r} = E r \, , \,\,\, \partial /\partial {\tilde r} 
= E^{-1} \partial / \partial r
\, , \,\,\, {\tilde{\bf p}} =
{\bf p} / E \,  .
\end{equation}
Dividing equation (\ref{eins}) by $E^{2}$ and setting $E^{2} = s$ for
convenience, one obtains:
\begin{equation}\label{48n}
\left\lbrack {\tilde{\bf p}} \bld{\alpha} + \frac{1}{2} m_{+}^{2}/s -
\frac{1}{2} + V({\tilde r})\, ( 1 - i {\tilde{\bf p}} (\bld{\alpha} \times
\bld{\sigma}_{2}) \, ) \right\rbrack \psi ({\tilde{\bf r}}) = 0 \, .
\end{equation}
Using the corresponding equation for $\chi^{\dagger}$, one arrives at:
\begin{eqnarray}\label{49n}
(s_{i}^{-1} - s_{j}^{-1} ) \int \chi_{i}^{\dagger}\,  m_{+}^{2}\, \psi_{j}
d^{3}{\tilde r}   & = & 0 \, , \quad s_{i} = E_{i}^{2}\, , \\
\label{49}
\int \chi_{i}^{\dagger}\,  
m_{+}^{2} \, \psi_{j} d^{3} {\tilde r} & = & \delta_{ij}
\, .
\end{eqnarray}
Remembering $m_{+}^{2}=m_{1}^{2} 
+ m_{2}^{2} + 2 m_{1}m_{2} \beta$, this is a simple
generalization of the static limit $m_{1}/m_{2} = 0$. For positronium, the
small components do not contribute to (\ref{49}) (a previously proposed
substitution $r = E \rho $ \cite{pil2} gives more complicated orthogonality
relations).

Equation (\ref{48n}) is explicitly CPT-invariant: Every bound state $s_{i}$ has
two different eigenvalues $E_{i}$, namely $E_{i} = \sqrt{s_{i}} \equiv 
m_{Ai}$ and $E_{i} = - \sqrt{s_{i}} \equiv - m_{Ai}$, 
where $m_{Ai}$ denotes the atomic mass in the state $i$ (an excited
atom is heavier than its ground state). The later value belongs 
to the antiatom of mass
$m_{{\bar A}i}$, i.e. $m_{{\bar A}i} \equiv m_{A i}$. This follows from the
CPT-invariance of QED, which ensures that the two-particle scattering amplitude
at negative $E$ describes the scattering amplitude of the two antiparticles.
The range of the dimensionless radial variable ${\tilde r}$ is $ 0 < {\tilde r}
< \infty $ both for atoms and for antiatoms. In the old variable $r$, antiatoms
have negative distances. This throws new light also on the static limit
$m_{1}/m_{2} = 0$. Here one defines $E_{e} = E - m_{2}$ as the electron
energy. For $E < 0$, one may use $E_{e} = E + m_{2}$. From the static Dirac
equation in the variable ${\tilde r}$, one obtains a spectrum which is
symmetric around $E = 0$ \cite{pil2}, for $V = - \alpha / {\tilde r}$. If one
wants to keep this symmetry as a result of CPT also in the case of a finite
nuclear charge distribution, one should parameterize $V({\tilde r})$ 
rather than
$V( r)$.

Of course, the mere CPT-invariance of a spectrum does not guarantee its
correctness.
Division of the Dirac-Breit equation by $E$ and
reformulation in terms of ${\tilde {\bf r}}_{1} = 
{\bf r}_{1} / E, \,\, {\tilde {\bf r}}_{2} = {\bf r}_{2} / E , \,\,
{\tilde {\bf r}} = {\bf r} / E$ also produces a CPT-invariant spectrum. But as
the interaction in this case does not reproduce the QED Born approximation at
all energies, one may hope that an 8-component formalism is again more
successful. The corresponding equation has been presented in (\ref{11n}) 
and the Born
approximation has been given in a suitable form in (\ref{34n}), but some
details are still missing. However, it is clear that the 8-component formalism
will be quadratic in the external potential $V_{e}$, but linear in $V$.

\section*{Acknowledgment}

This work has been supported by the {\sl Deutsche Forschungsgemeinschaft}.

\begin{appendix}

\section{Combinations of the hyperfine operator with $\bld{\alpha} 
 \lowercase{{\bf p}}  $} 
\label{appa}

Writing $\bld{\sigma}_{1}{\bf p}$ as $\frac{1}{2} ( \bld{\sigma} + \Delta
\bld{\sigma} ) {\bf p}$, one observes from (\ref{22n}) that $\Delta
\bld{\sigma}$ and $\bld{\sigma}^{\times}$ transform triplets $\chi_{t}$ into
the singlet $\chi_{s}$ and vice versa. As a result, the combination required in
(\ref{fuenf}) may be written as:
\begin{equation}\label{a1}
\frac{1}{2} {\bf p} \Delta \bld{\sigma} - i V {\bf p} \bld{\sigma}^{\times} / E
= \left\lbrack \frac{1}{2} + ( P_{t} - P_{s} ) V / E \right\rbrack {\bf p}
\Delta\bld{\sigma} \, .
\end{equation}
For total angular momentum $f$, the triplet states with $l=f$ are excluded
from ${\bf p} \Delta \bld{\sigma}$ by parity conservation. Thus one has in the
notation of \cite{mal}:
\begin{equation}\label{a2}
{\tilde {\bf p}} \Delta \bld{\sigma} = 2 i \left( \begin{array}{cccc}
0 & 0 & 0 & \partial_{-}\\
0 & 0 & 0 & - F /r \\
0 & 0 & 0 & 0 \\
\partial_{+} & F/r & 0 & 0 \end{array} \right) \, ,\quad
{\tilde{\bf p}} \bld{\sigma}^{\times} = 2 \left(
\begin{array}{cccc}
0 & 0 & 0 & - \partial_{-} \\
0 & 0 & 0 & F/r \\
0 & 0 & 0 & 0 \\
\partial_{+} & F/r & 0 & 0 
\end{array} \right) \, ,
\end{equation}
with ${\tilde {\bf p}} =r {\bf p} / r\, , \,\, \partial_{\pm} = \partial_{r}
\pm 1 / r$ and $ F = \sqrt{f(f+1)}$.

\section{Connection with the form derived from Breit operators}\label{appb}

We substitute in (\ref{fuenf}) $\psi = e^{x}\, \psi_{B}$ and multiply the
equation by $e^{-x}$ from the left, where the operator $x$ is of the order of
$V/E$ and commutes with $\gamma_{5}$ and $\beta$. To order $\alpha^{4}$, one
may then approximate:
\begin{equation}\label{B1}
e^{-x} \, {\bf p}\bld{\alpha}\,  
e^{x} \approx (1-x)\, {\bf p} \bld{\alpha}\,  (1 + x)
\approx {\bf p}\bld{\alpha} + \lbrack {\bf p}\bld{\alpha}, x \rbrack 
\, , \,\,\,
e^{-x}\,  K_{I} / E \, e^{x} \approx K_{I}/ E \, .
\end{equation}
Choosing now
\begin{equation}\label{B2}
x= - \bld{\sigma}_{1}\bld{\sigma}_{2} V / 2 E
\end{equation}
and extracting a common factor $\gamma_{5}$, one has:
\begin{equation}\label{B3}
e^{-x} \, ( {\bf p} \bld{\sigma}_{1} - i V {\bf p} 
\bld{\sigma}^{\times} / E )\,
e^{x} = {\bf p} \bld{\sigma}_{1} - \frac{1}{2E} ( i  
\bld{\sigma}^{\times}
- \bld{\sigma}_{2} ) \lbrack V , {\bf p} \rbrack \, .
\end{equation}
The second piece is the hyperfine operator derived from Breit operators
\cite{mal} and is known to reproduce the hyperfine structure of leptonium to
order $\alpha^{4}$, including positronium \cite{pil2}. It can be rewritten in
compact form:
\begin{equation}\label{B4}
- \frac{1}{2E} ( \bld{\sigma}^{\times} + i \bld{\sigma}_{2} )
\lbrack V , \bld{\nabla} \rbrack = \frac{i}{2 E} \lbrack \bld{\sigma}_{1}
\bld{\nabla}  , V
\rbrack \bld{\sigma}_{1} \bld{\sigma}_{2}\, .
\end{equation}
%
 
\section{Spin summation}\label{appc}

The propagator of a lepton-antilepton pair will be needed in the perturbative
interaction with radiation. Here we merely perform the spin summation for the
trivial case of a free pair. We remind the reader that in the 16-component
formalism, one defines for particles $i=1,2$:
\begin{equation}\label{c1}
\gamma_{i}^{0} = \beta_{i}\, , \quad \bld{\gamma}_{i} = \gamma_{i}^{0}
\bld{\alpha}_{i} = \gamma_{i}^{0} \gamma_{5i} \bld{\sigma}_{i} \, ,\quad 
/\!\!\!p_{i} =  p_{i} \gamma_{i} \, ,
\end{equation}
which leads to the following form of the spin summation:
\begin{equation}\label{c2}
\sum_{{\mbox{\tiny spins}}} u_{1} u_{2} {\bar u}_{1} {\bar u}_{2} = (
/\!\!\!p_{1} + m_{1} ) ( /\!\!\!p_{2} + m_{2} ) \, .
\end{equation}
A similar notation may also be used in the 8-component version, but with the
understanding $\gamma_{1}^{0} = \gamma_{2}^{0} = \beta\, ,\,\, \gamma_{51} =
\gamma_{52} = \gamma_{5}\, .$
Consequently, $\, /\!\!\!p_{1} /\!\!\!p_{2} 
\not= /\!\!\!p_{2} /\!\!\!p_{1}$, but:
\begin{equation}\label{c3}
/\!\!\!p_{1} \beta /\!\!\!p_{2} = /\!\!\!p_{2} \beta /\!\!\!p_{1}\, , \quad
/\!\!\!p_{1} /\!\!\!p_{2} \beta = \beta /\!\!\!p_{2} /\!\!\!p_{1} \, .
\end{equation}
The free leptonium equation (\ref{11n}) and the corresponding equation for
$w^{+}$ become:
\begin{equation}\label{c4}
( m_{2} \beta /\!\!\!p_{1} - m_{1} /\!\!\!p_{2} ) u = 0 \, , \quad w^{+} (
m_{2} \beta /\!\!\!p_{1} - m_{1} /\!\!\!p_{2} ) = 0 \, .
\end{equation}
From (\ref{c3}) and (\ref{c4}), one easily verifies the following spin
summation:
\begin{equation}
s = \sum_{{\mbox{\tiny spins}}} u w^{\dagger} = m_{2} /\!\!\!p_{1} \beta +
m_{1} /\!\!\!p_{2}\, .
\end{equation}
It is remarkable that this $8\times 8$-matrix is linear in $/\!\!\!p_{1}$ and
$/\!\!\!p_{2}$, while the $16 \times 16$-matrix (\ref{c2}) also contains
$/\!\!\!p_{1} /\!\!\!p_{2}$.

\end{appendix}


\end{document}